\documentclass[preprint, 5p, twocolumn]{elsarticle}

\usepackage{lineno,hyperref}
\usepackage{lineno}
\usepackage{graphicx}
\usepackage{subcaption}
\usepackage{amsmath}
\usepackage{verbatim}
\usepackage{amssymb}
\usepackage{color}
\usepackage{xcolor}

\usepackage{mathrsfs}
\modulolinenumbers[5]

\journal{Journal of \LaTeX\ Templates}

\begin{document}

\begin{frontmatter}

\title{Normal modes for two-dimensional gravitating kinks}


\author[SoP,ITP]{Yuan Zhong\corref{mycorrespondingauthor}}
\cortext[mycorrespondingauthor]{Corresponding author}
\ead{zhongy@mail.xjtu.edu.cn}
\address[SoP]{School of Physics, Xi'an Jiaotong University, Xi'an 710049, China}
\address[ITP]{Institute of Theoretical Physics, Xi'an Jiaotong University, Xi'an 710049, China}

\begin{abstract}
We study small perturbations around an arbitrary static kink solution of a two-dimensional (2D) gravity-scalar system, where the gravity part is described by a subclass of 2D dilaton gravity theory, and the scalar matter field has generalized dynamics. We expand the action around an arbitrary static solution and keep terms up to the second order of the perturbations. After variation the linear-order action leads to background field equations, as expected. The quadratic action of the normal modes are obtained after fixing the gauge and using the constraint equation. The linear perturbation equations obtained from the quadratic action are consistent with those obtained by linearizing the field equations under the dilaton gauge. All the calculations are assisted by a Mathematica code, which is also provided as a supplementary material.
\end{abstract}


\end{frontmatter}



\section{Introduction}
Two-dimensional (2D) gravity theories are simpler mathematically, but still reflect part of the features of their higher-dimensional counterparts~\cite{Jackiw1985,Teitelboim1983}. For this reason, 2D gravity theories have been applied to discuss quantization of gravity~\cite{Henneaux1985,Alwis1992}, gravitational collapse~\cite{VazWitten1994,VazWitten1996}, black hole evaporation~\cite{CallanGiddingsHarveyStrominger1992,BilalCallan1993,RussoSusskindThorlacius1992,RussoSusskindThorlacius1992a,RussoSusskindThorlacius1993},  and many other difficult issues, see~\cite{Brown1988,Thorlacius1995,NojiriOdintsov2001d,GrumillerKummerVassilevich2002} for comprehensive reviews on early works. Recently, the study of the Sachdev-Ye-Kitaev (SYK) model \cite{SachdevYe1993,Kitaev2015} displays a potential application of 2D gravity in the study of gauge/gravity duality~\cite{AlmheiriPolchinski2015,MaldacenaStanfordYang2016,MaldacenaStanford2016,Jensen2016},  see~\cite{Rosenhaus2019,Sarosi2018,Trunin2021} for pedagogical introductions.

In parallel to the study of 2D black holes, it is natural to consider 2D gravity models with nonlinear scalar fields, and regarding smooth gravitating kink solutions as 2D versions of some 5D thick brane solutions. The later have been extensively studied in the past two decades~\cite{SkenderisTownsend1999,DeWolfeFreedmanGubserKarch2000,Gremm2000,DzhunushalievFolomeevMinamitsuji2010,Liu2018}  as they offer some regular extensions of the Randall-Sundrum thin brane world models~\cite{RandallSundrum1999,RandallSundrum1999a,GoldbergerWise1999,GoldbergerWise1999a}.

Although the study of 2D self-gravitating kinks can be traced back to 1995~\cite{ShinSoh1995,Stoetzel1995,JohngShinSoh1995,JohngShinSoh1996,AlvesBazeiaBezerra1999}, most of these solutions contain some space-time singularities. To describe 2D thick brane, we require some gravitating kink solutions without any singularity.

To the author's knowledge, the first solution of this type was reported by St\"otzel~\cite{Stoetzel1995}, who considered the 2D gravity model of Mann, Morsink, Sikkema and Steele (MMSS)~\cite{MannMorsinkSikkemaSteele1991}. St\"otzel's solution has no singularity, and the corresponding metric is asymptotic AdS$_2$, and therefore, is an excellent example of 2D thick brane.

In a recent work~\cite{Zhong2021}, the present author pointed out that the procedure used by St\"otzel for deriving his solution is nothing but the superpotential method, which has been used repeatedly in the study of 5D thick branes~\cite{SkenderisTownsend1999,DeWolfeFreedmanGubserKarch2000,BazeiaFurtadoGomes2004,BazeiaGomes2004,BazeiaGomesLosanoMenezes2009,AfonsoBazeiaLosano2006,BazeiaLobLosanoMenezes2013,ZhongLiuZhao2014a}.  This observation makes it an easy task to construct analytical solutions of 2D gravitating kink, some examples can be found in Ref.~\cite{Zhong2021} for the case with canonical scalar field, and~\cite{ZhongLiLiu2021} for the case with noncanonical scalar field.  All these solutions are regular, and can be regarded as 2D thick branes.

Before using these solutions for further explorations, one should carefully discuss the stability of these solutions under small field perturbations. This issue was first analyzed in Ref.~\cite{Zhong2021}, where the author conducted a direct linearization of the field equations. After gauging out the perturbation of the dilaton field, and using the constraint equations, a Schr\"odinger-like equation is obtained for the physical perturbation. In particular, the Hamiltonian operator of the Schr\"odinger-like equation can be factorized as the form $\hat{H}=\hat{\mathcal{A}}\hat{\mathcal{A}}^\dagger $, which indicates that the eigenvalues are positive-semidefinite, and therefore, the solutions are stable. Similar results were also obtained for the case with noncanonical scalar field~\cite{ZhongLiLiu2021}.  Remarkablly, the perturbation equations derived in Refs.~\cite{Zhong2021,ZhongLiLiu2021} take similar form as the scalar perturbation equations of some 5D thick brane model with Einstein gravity~\cite{Giovannini2001a,Giovannini2003,ZhongLiu2013}. This verified the conjecture that the MMSS model may be the closest thing there is to general relativity in two dimensions~\cite{MannRoss1993}.

In the present work, we notice that there are two different approaches for deriving the linear perturbation equations. In the first approach, one directly linearizes the field equations, as has been done in Refs.~\cite{Zhong2021,ZhongLiLiu2021}. In the other, one first expands the action around a background solution, and keep only the quadratic terms of the perturbations. After canonicalizing this quadratic action, one can easily read out the master equations for the normal  modes of the perturbations. Naively, one may expect that these two  approaches leads to equivalent results. However, counterexamples have already been reported in some 5D thick brane models~\cite{Giovannini2001a,Giovannini2003,ZhongLiu2013}.

In brief, Refs.~\cite{Giovannini2001a,Giovannini2003,ZhongLiu2013} tell us that one must fix the gauge properly if he adopt the equation of motion approach. Otherwise, the Schr\"odinger-like equation might be different from the one obtained from the quadratic approach. For example, if one starts with the widely used longitude gauge, he would obtain a factorizable Hamiltonian operator $\hat{H}_-=\hat{\mathcal{A}}^\dagger\hat{\mathcal{A}} $ for the physical perturbation modes. While the Hamiltonian operator derived from the quadratic action approach is $\hat{H}=\hat{\mathcal{A}} \hat{\mathcal{A}}^\dagger$. One may recognize that $\hat{H}$ and $\hat{H}_-$ are a pair of superpartners in the sense of supersymmetric quantum mechanics~\cite{Witten1981,CooperKhareSukhatme1995}.

Usually, a pair of superpartner Hamiltonians share the same spectra, except their zero modes~\cite{CooperKhareSukhatme1995}. So the properties of $\hat{H}$ can be learnt from $\hat{H}_-$, and vise versa. The situation changes, however, if the potential term of either $\hat{H}$ or $\hat{H}_-$ diverges. In this case, the supersymmetry might be explicitly broken, and the eigenvalues  of $\hat{H}$ or $\hat{H}_-$ are only partial degenerated, if the potential is symmetric, or completely non-degenerated, if the potential is asymmetric~\cite{CooperKhareSukhatme1995,JevickiRodrigues1984,RoyRoychoudhury1985,CasahorranNam1991,PanigrahiSukhatme1993}. Even worse, one of the superpartner Hamiltonians might has a eigenstate with negative eigenvalue, a sign for instability! Therefore, when singular potential appears, one must adopt the quadratic approach, as it gives the correct Hamiltonian for the normal perturbation modes.

Unfortunately, for 2D gravitating kink solutions considered in Refs.~\cite{Zhong2021,ZhongLiLiu2021}, singular potentials in the Schr\"odinger-like equation seems unavoidable within some parameter ranges. For this reason, it is necessary to conduct an independent derivation of the linear perturbation equations from the quadratic action approach, and see if they are equivalent with those obtained in Refs.~\cite{Zhong2021,ZhongLiLiu2021}.

In the next section, we briefly review the main results of gravitational perturbation theory for an arbitrary background metric solution of the following model~\cite{ZhongLiLiu2021}:
\begin{eqnarray}\label{action}
S=\frac{1}{\kappa} \int d^{2} x \sqrt{-g}\left[\mathcal{X}+\varphi R+\kappa \mathcal{L}(\phi,X)\right],
\end{eqnarray}
where $\kappa$ is the gravitational coupling constant, $g$ is the determinant of the metric, $\mathcal{X}\equiv -\frac{1}{2} g^{\mu\nu}\nabla_{\mu} \varphi \nabla_{\mu} \varphi$ and $X\equiv -\frac{1}{2} g^{\mu\nu}\nabla_{\mu} \phi \nabla_{\mu} \phi$ are the kinetic terms of the dilaton field and the scalar field, respectively.
In Sec.~\ref{SectionThree}, we come back to the case with static solutions. After taking the dilaton gauge, we derive the quadratic action of the normal perturbation modes. All our calculations are conducted with a Mathematica code, which can be found in the supplementary material. The results are summarized in Sec.~\ref{sec_con}.

\section{Gravitational perturbation theory: a brief review}
Consider a small perturbation $\delta g_{\mu\nu}$ around an arbitrary background metric solution $g_{\mu\nu}$.
If we denote the perturbation of the inverse metric as $\delta g^{\nu\sigma}$, and require the orthogonal relation
\begin{eqnarray}
(g_{\mu\nu}+\delta g_{\mu\nu})(g^{\nu\sigma}+\delta g^{\nu\sigma})=\delta^\sigma_\mu,
\end{eqnarray}
then we would obtain the following equation:
\begin{eqnarray}
g_{\mu\nu}\delta g^{\nu\sigma}+g^{\nu\sigma}\delta g_{\mu\nu}+\delta g_{\mu\nu}\delta g^{\nu\sigma}=0.
\end{eqnarray}
Obviously, the last two term are not of the same order, therefore, this equation can only be satisfied order by order, which means that $\delta g^{\nu\sigma}=\delta^{(1)} g^{\nu\sigma}+\delta^{(2)} g^{\nu\sigma}+\cdots$ is a summation of perturbations with different orders.

To the first and second orders, one obntains
\begin{eqnarray}
\delta^{(1)} g^{\mu\nu}&=&-g^{\mu\rho}g^{\nu\sigma}\delta g_{\rho\sigma},\\
\delta^{(2)} g^{\mu\nu}&=&-g^{\mu\rho}\delta g_{\rho\sigma}\delta^{(1)}  g^{\sigma\nu},
\end{eqnarray}
respectively. The Christoffel symbol is completely defined by the metric, and its first and second-order perturbations take the following forms:
\begin{eqnarray}
\delta^{(1)}\Gamma_{\mu \nu}^{\lambda}&=&\frac{1}{2} \delta^{(1)}g^{\lambda \sigma}\left(\partial_{\mu} g_{\sigma\nu }+\partial_{\nu} g_{\sigma \mu}-\partial_{\sigma} g_{\mu \nu}\right)\nonumber\\
&+&\frac{1}{2} g^{\lambda \sigma}\left(\partial_{\mu} \delta g_{\sigma\nu }+\partial_{\nu} \delta g_{\sigma \mu}-\partial_{\sigma} \delta g_{\mu \nu}\right),\\
\delta^{(2)}\Gamma_{\mu \nu}^{\lambda}&=&\frac{1}{2} \delta^{(2)}g^{\lambda \sigma}\left(\partial_{\mu} g_{\sigma\nu }+\partial_{\nu} g_{\sigma \mu}-\partial_{\sigma} g_{\mu \nu}\right)\nonumber\\
&+&\frac{1}{2} \delta^{(1)}g^{\lambda \sigma}\left(\partial_{\mu} \delta g_{\sigma\nu }+\partial_{\nu} \delta g_{\sigma \mu}-\partial_{\sigma} \delta g_{\mu \nu}\right).
\end{eqnarray}
Similarly, one can write out the perturbations of the Ricci tensor
\begin{eqnarray}
\delta^{(1)} R_{\mu \nu}&=&\partial_{\rho} \delta^{(1)}\Gamma_{\mu\nu}^{\rho}-\partial_{\nu}\delta^{(1)} \Gamma_{\mu \rho}^{\rho}
+ \delta^{(1)}\Gamma_{\lambda \rho}^{\rho} \Gamma_{\mu \nu }^{\lambda}\nonumber\\
&+&\Gamma_{\lambda \rho}^{\rho}  \delta^{(1)}\Gamma_{\mu \nu }^{\lambda}
- \delta^{(1)}\Gamma_{\lambda \nu }^{\rho} \Gamma_{\mu \rho}^{\lambda}\nonumber\\
&-&\Gamma_{\lambda \nu }^{\rho}  \delta^{(1)}\Gamma_{\mu \rho}^{\lambda},\\
\delta^{(2)} R_{\mu \nu}&=&\partial_{\rho} \delta^{(2)}\Gamma_{\mu\nu}^{\rho}
-\partial_{\nu}\delta^{(2)} \Gamma_{\mu \rho}^{\rho}
+ \delta^{(2)}\Gamma_{\lambda \rho}^{\rho} \Gamma_{\mu \nu }^{\lambda}\nonumber\\
&+&\delta^{(1)}\Gamma_{\lambda \rho}^{\rho}  \delta^{(1)}\Gamma_{\mu \nu }^{\lambda}
+ \Gamma_{\lambda \rho}^{\rho} \delta^{(2)}\Gamma_{\mu \nu }^{\lambda}\nonumber\\
&-&
 \delta^{(2)}\Gamma_{\lambda \nu }^{\rho} \Gamma_{\mu \rho}^{\lambda}
-\delta^{(1)}\Gamma_{\lambda \nu }^{\rho}  \delta^{(1)}\Gamma_{\mu \rho}^{\lambda}\nonumber\\
&-&\Gamma_{\lambda \nu }^{\rho}  \delta^{(2)}\Gamma_{\mu \rho}^{\lambda},
\end{eqnarray}
and the scalar curvature
\begin{eqnarray}
\delta^{(1)} R &=&\delta^{(1)}g^{\mu \nu}R_{\mu \nu}+g^{\mu \nu} \delta^{(1)} R_{\mu \nu},\\
\delta^{(2)} R &=&\delta^{(2)}g^{\mu \nu}R_{\mu \nu}
+\delta^{(1)}g^{\mu \nu} \delta^{(1)} R_{\mu \nu}\nonumber\\
&+&g^{\mu \nu} \delta^{(2)} R_{\mu \nu}.
\end{eqnarray}
The perturbations of $\sqrt{- g}$ are also useful
\begin{eqnarray}
\delta^{(1)} \sqrt{-g}&=&\frac{1}{2} \sqrt{-g} g^{\mu\nu}\delta g_{\mu\nu}, \\
\delta^{(2)} \sqrt{-g}&=&\frac{1}{8} \sqrt{-g}\left[(g^{\mu\nu}\delta g_{\mu\nu})^{2}
+2 \delta g_{\mu\nu} \delta^{(1)} g^{\mu\nu}\right].
\end{eqnarray}

In addition to the metric perturbation, we also have perturbations of the dilaton and the scalar fields, which are denoted as $\delta\varphi$ and $\delta\phi$, respectively. One can testifies that the perturbations of the dilaton kinetic term are
\begin{eqnarray}
\delta^{(1)}\mathcal{X}&=&-\frac12\delta^{(1)}g^{\mu\nu}\nabla_\mu\varphi \nabla_\nu\varphi
-g^{\mu\nu}\nabla_\mu\delta\varphi \nabla_\nu\varphi,\\
\delta^{(2)}\mathcal{X}&=&-\frac12\delta^{(2)}g^{\mu\nu}\nabla_\mu\varphi \nabla_\nu\varphi
-\delta^{(1)}g^{\mu\nu}\nabla_\mu\delta\varphi \nabla_\nu\varphi\nonumber\\
&-&\frac12g^{\mu\nu}\nabla_\mu\delta\varphi \nabla_\nu\delta\varphi,
\end{eqnarray}
which after replacing $\varphi\to \phi$ give us the expressions of $\delta^{(1)}X$ and $\delta^{(2)}X$.

Finally, the perturbations of the scalar Lagrangian density are
\begin{eqnarray}
\delta^{(1)}\mathcal{L}&=&\mathcal{L}_\phi\delta\phi+\mathcal{L}_X\delta^{(1)}X,\\
\delta^{(2)}\mathcal{L}&=&\mathcal{L}_{X\phi}\delta\phi \delta^{(1)}X
+\frac12\mathcal{L}_{\phi\phi}(\delta\phi)^2\nonumber\\
&+&\frac12\mathcal{L}_{XX}(\delta^{(1)}X)^2
+\mathcal{L}_{X}\delta^{(2)}X.
\end{eqnarray}
In this work, we defined $\mathcal{L}_\phi\equiv \frac{\partial \mathcal{L}}{\partial \phi}$, $\mathcal{L}_X\equiv \frac{\partial \mathcal{L}}{\partial X}$, and so on.

With all these preparation, we can now write out the perturbations of the action \eqref{action}:
\begin{eqnarray}
&&\delta^{(1)}S=\frac 1\kappa \int d^2x \{\delta^{(1)}\sqrt{-g}[\mathcal{X}+\varphi R+\kappa \mathcal{L}]
\nonumber\\
&+&\sqrt{-g}[\delta^{(1)}\mathcal{X}+\delta\varphi R+\varphi \delta^{(1)}R+\kappa \delta^{(1)}\mathcal{L}]\},\end{eqnarray}
and
\begin{eqnarray}
&&\delta^{(2)}S=\frac 1\kappa \int d^2x \{ \delta^{(2)}\sqrt{-g}[\mathcal{X}+\varphi R+\kappa \mathcal{L}]\nonumber\\
&+&\delta^{(1)}\sqrt{-g}[\delta^{(1)}\mathcal{X}+\delta\varphi R+\varphi \delta^{(1)}R+\kappa \delta^{(1)}\mathcal{L}]\nonumber\\
&+&\sqrt{-g}[\delta^{(2)}\mathcal{X}+\delta\varphi \delta^{(1)} R
+\varphi \delta^{(2)}R
+\kappa \delta^{(2)}\mathcal{L}]\}.
\end{eqnarray}

No doubt, it would be very complicate to calculate the perturbations of the action manually. Fortunately, in two dimension these calculations can be accomplished by writing Mathematica codes, see the supplementary material.

\section{The case of static solution}
\label{SectionThree}
Now, let us turn to the special case with static solutions, where $g_{\mu\nu}=g_{\mu\nu}(x)$, $\varphi=\varphi(x)$, $\phi=\phi(x)$. In two dimensions one can always express a static metric into the following form:
\begin{eqnarray}
\label{metricXCord}
  ds^2=-e^{2A(x)}dt^2+dx^2.
\end{eqnarray}
This coordinate system is very convenient for constructing first-order formalism, from which analytical gravitating kinks can be derived, see Refs.~\cite{Zhong2021,ZhongLiLiu2021} for details.

However, to discuss the perturbation issue, it is more convenient to introduce a spatial transformation
\begin{eqnarray}
r\equiv\int e^{-A(x)}dx,
\end{eqnarray}
with which the metric can be written as a conformally flat form
\begin{eqnarray}
\label{Gmn}
g_{\mu\nu}(r)=e^{2A(r)}\eta_{\mu\nu}.
\end{eqnarray}
Following Refs.~\cite{Zhong2021,ZhongLiLiu2021}, we define the metric  perturbation as
\begin{eqnarray}
\label{delGmn}
\delta g_{\mu\nu}(t,r)&\equiv& e^{2A(r)} h_{\mu\nu}(t,r)\nonumber\\
&=&e^{2A(r)} \left(
\begin{array}{cc}
 h_{00}(t,r) & \Phi (t,r) \\
 \Phi (t,r) & h_{rr}(t,r) \\
\end{array}
\right).
\end{eqnarray}
Note that although the background solution is static, the field perturbations $\delta g_{\mu\nu}(t,r)$, $\delta \varphi(t,r)$, $\delta \phi(t,r)$ are all functions of both $t$ and $r$.

Using Eqs.~\eqref{Gmn} and \eqref{delGmn}, one can proof that
\begin{eqnarray}
&&\delta^{(1)}S=\frac1\kappa\int d^2x\big\{
-2 \varphi  A'  \dot{\Phi }
-2 \varphi  \dot{\Phi }'
 \nonumber\\
&- &\varphi '  \delta \varphi '
-2  A'' \delta \varphi
+  \kappa e^{2 A}  \mathcal{L} _\phi  \delta \phi
-\kappa    \mathcal{L} _X \phi ' \delta \phi '
\nonumber\\
&+&\frac{1}{4} h_{00} \left(4 \varphi  A''-2 e^{2 A} \kappa  \mathcal{L}
+{\varphi ' }^2\right) \nonumber\\
&+&\varphi  A' h_{00}'
+\varphi  h_{00}''
+\varphi  A' h_{rr}'
+\varphi  \ddot{h}_{rr}
\nonumber\\
&+&\frac{1}{4} h_{rr} \left(4 \varphi  A''
+2 e^{2 A} \kappa  \mathcal{L}
+2 \kappa  \mathcal{L} _X {\phi ' }^2
+{\varphi '}^2\right)
\big\}.
\end{eqnarray}
Here we used over dots and primes to denote the derivatives with respect to $t$ and $r$, respectively.
Obviously, the $\Phi$ terms can be written as a total derivative term (as $\dot{A}=0=\dot{\varphi}$), and does not lead to field equation.

After partial integrations and deminding the coefficient of $\delta\varphi$ equals to zero, one immediately obtains the dilaton equation
\begin{eqnarray}
\label{eqDilaton}
\varphi=2A.
\end{eqnarray}
Similarly, for $\delta\phi$, $h_{00}$ and $h_{rr}$ terms, the corresponding field equations are:
\begin{eqnarray}
\label{EqEOM}
(\mathcal{L}_X \phi '  )^\prime + e^{2 A}  \mathcal{L}_\phi =0,
\end{eqnarray}
\begin{eqnarray}
\label{EqEE1}
4 A'' -2 A'^2 =\kappa  e^{2 A} \mathcal{L},
\end{eqnarray}
and
\begin{eqnarray}
\label{EqEE2}
2 {A' }^2=\kappa  e^{2 A} \mathcal{L}+\kappa  \mathcal{L}_X {\phi ' }^2,
\end{eqnarray}
where we have used the dilaton equation \eqref{eqDilaton}.
The combination of Eqs.~\eqref{EqEE1} and \eqref{EqEE2} also gives another useful equation
\begin{eqnarray}
\label{EqEE3}
-4 A'' +4 {A' }^2= \kappa  \mathcal{L}_X {\phi ' }^2.
\end{eqnarray}
The above field equations are consistent with those obtained in Ref.~\cite{ZhongLiLiu2021}.

The quadratic action of the perturbations are much more complicated. To simplify the calculation, we first note that the general invariance of action \eqref{action} under the general coordinate transformation \begin{eqnarray}
x^\mu \to \tilde{x}^\mu=x^\mu+\xi^\mu(t,r),
  \end{eqnarray}
induces an invariance of $\delta^{(2)}S$ under the following gauge transformations~\cite{Zhong2021}:
\begin{eqnarray}
\Delta h_{00}&=&2\partial _t\xi ^0+2\frac{a'}{a}\xi ^1,
\\
\Delta \Phi&=&-\partial _t\xi ^1+\partial _r\xi ^0,
\\
\Delta h_{11}&=&-2\partial _r\xi ^1-2\frac{a'}{a}\xi ^1,\\
 \Delta \delta \phi&=&-\phi^{\prime} \xi^{1} ,\\
  \Delta \delta \varphi&=&-\varphi^{\prime} \xi^{1} ,
  \end{eqnarray}
where $\Delta h_{00} \equiv \widetilde{h}_{00}-h_{00}$, and so on.

To fix the gauge degrees of freedom,  we first define a new variable $\Xi \equiv  2\dot{\Phi}- h_{00}^{\prime}$, whose gauge transformation only depends on $\xi^1$:
\begin{eqnarray}
\Delta \Xi=-2\left[\ddot{\xi}^{1}+\left(A^{\prime} \xi^{1}\right)^{\prime}\right].
  \end{eqnarray}
Thus, the introducing of $\Xi$ eliminates the the degree of freedom of $\xi^0$. To fix the other, we simply chose the gauge condition such that $\delta \varphi=0$ \cite{Zhong2021}.

After this gauge fixing process, we can separate $\delta^{(2)}S$ into three parts
\begin{eqnarray}
\delta^{(2)}S=\delta^{(2)}S_\Xi+\delta^{(2)}S_{h_{rr}}+\delta^{(2)}S_{\delta\phi},
\end{eqnarray}
where
\begin{itemize}
\item $\delta^{(2)}S_{\Xi}\equiv$ all the terms with $\Xi$ and its derivatives.
\item $\delta^{(2)}S_{h_{rr}}\equiv$ all the terms with $h_{rr}$ and its derivatives, but without $\Xi$ and its derivatives.
\item $\delta^{(2)}S_{\delta\phi}\equiv$ all the terms with only $\delta\phi$ and its derivatives, in other words, the quadratic terms of $\delta\phi$.
\end{itemize}
The above definitions prevent us from overcounting the cross terms.

By conducting partial integrations many times and discarding all the boundary terms, we obtain
\begin{eqnarray}
\delta^{(2)}S_{\Xi}&=&\frac{1}{\kappa}\int d^2x\left\{\frac12\kappa \phi ' \mathcal{L}_X \delta \phi
- A' h_{rr} \right\} \Xi ,
\end{eqnarray}
\begin{eqnarray}
\delta^{(2)}S_{h_{rr}}&=&\frac{1}{\kappa}\int d^2x\bigg\{
\frac{1}{2} \gamma  \left(A''-A'^2\right)h_{rr}^2 \nonumber\\
&+& \frac{1}{2}A'^2 h_{rr}^2
+2 \gamma   \left(A'^2-A''\right)\frac{ \delta \phi ' }{\phi '}h_{rr}
\nonumber\\
&+&2  A' \left(A''-A'^2\right)\frac{\delta \phi }{\phi '} h_{rr}\nonumber\\
&+&\gamma   \left(A''-A'^2\right)\frac{X'  }{X } \frac{\delta \phi }{\phi '}h_{rr}\bigg\},
\end{eqnarray} where
\begin{eqnarray}
\gamma\equiv 1+2\frac{\mathcal{L}_{XX}X }{\mathcal{L}_{X} },
\end{eqnarray}
and
\begin{eqnarray}
\delta^{(2)}S_{\delta \phi}&=&\frac{1}{\kappa}\int d^2x\bigg\{
\frac{1}{2} \kappa  e^{-2 A}   \mathcal{L}_{XX} \phi'^2 \delta \phi'^2
\nonumber\\
&+&\frac{1}{2}  \kappa  e^{2 A} \mathcal{L}_{\phi\phi}   \delta \phi ^2
-\frac{1}{2} \kappa  \mathcal{L}_{X}  \delta \phi'^2
 \nonumber\\
&+&\frac{1}{2}  \kappa  \mathcal{L}_{X} \dot{\delta \phi }^2
-\kappa  \mathcal{L}_{X\phi} \phi '  \delta \phi   \delta \phi '  \bigg\}.
\end{eqnarray}

Obviously, $\Xi$ is a Lagrangian multiplier, whose variation leads to a constraint equation:
\begin{eqnarray}
\label{ConsEq1}
2 A' {h}_{rr}=\kappa  \mathcal{L}_X \phi ' {\delta \phi }.
\end{eqnarray}
The same equation was also derived in Ref.~\cite{ZhongLiLiu2021} from the equation of motion approach.

After eliminating $h_{rr}$, $\delta^{(2)}S$ becomes a quadratic action of $\delta\phi$. The simplification of $\delta^{(2)}S$ is the most intricate part of the present work. A key observation is that all the coefficients of can be rearranged as functions of only $A', \phi'$, $\gamma$ (or equivalently $\mathcal{L}_{XX}$), $X$, $\mathcal{L}_X$, and the derivatives of $\gamma$, $X$, and $\mathcal{L}_X$.

To see this clearly, we first note that Eq.~\eqref{EqEE3} allows us to express $A''$ as a function of  $A', \phi'$, and $\mathcal{L}_X$:
\begin{eqnarray}
\label{EqApp}
A''=A'^2-\frac{1}{4} \kappa   \mathcal{L}_X \phi'^2.
\end{eqnarray}
Similarly, by taking the derivative of $X$, we can express $\phi '' $ in terms of  $A', \phi'$, $X$ and $X'$
\begin{eqnarray}
\label{EqPhipp}
\phi ''= A' \phi '+\frac12 \frac{X' }{ X}\phi '.
\end{eqnarray}
Then, by taking the derivative of $\mathcal{L}_X$ we obtain
\begin{eqnarray}
\label{EqLXphi}
\mathcal{L}_{X\phi }= \frac{1}{\phi '}\left(\mathcal{L}_X' -\mathcal{L}_{XX} X' \right).
\end{eqnarray}
Finally, by taking the derivative of Eq.~\eqref{EqEOM}, and using Eqs.~\eqref{EqApp}-\eqref{EqLXphi}, we get
\begin{eqnarray}
\label{EqLphph}
\mathcal{L}_{\phi \phi }&=& \frac{\mathcal{L}_X X''}{\phi'^2}
+\frac{X'^2 \mathcal{L}_{{XX}}}{\phi'^2}
-\frac{\mathcal{L}_X X'^2}{2 X \phi'^2}\nonumber\\
&+&\frac{X' \mathcal{L}_X'}{\phi'^2}
-\frac{1}{2} \kappa  X \mathcal{L}_X^2
+\frac{2 X \mathcal{L}_X''}{\phi'^2}.
 \end{eqnarray}
With Eqs.~\eqref{EqApp}-\eqref{EqLphph}, a direct calculation shows that $\delta^{(2)}S$ takes the following form:
\begin{eqnarray}
\delta^{(2)}S=\frac{1}{2} \int d^2x  \mathcal{L}_X \left\{
- \ddot{\delta \phi }
+ \gamma  \delta \phi ''
+\gamma U \delta \phi
 \right\}\delta \phi,
\end{eqnarray}
where
\begin{eqnarray}
U&\equiv& -\frac{\kappa  \mathcal{L}_X X' \phi'^2}{2 X A'}
-\frac{\kappa  \gamma ' \mathcal{L}_X \phi'^2}{4 \gamma  A'}
-\frac{\kappa ^2 \mathcal{L}_X^2 \phi'^4}{8 A'^2}
\nonumber\\
&-&\frac{\kappa  \phi'^2 \mathcal{L}_X'}{2 A'}
-\frac{\gamma ''}{2 \gamma }-\frac{X''}{2 X}
-\frac{\gamma ' X'}{2 \gamma  X}
+\frac{X'^2}{4 X^2}\nonumber\\
&-&\frac{X' \mathcal{L}_X'}{2 X \mathcal{L}_X}
-\frac{\gamma ' \mathcal{L}_X'}{\gamma  \mathcal{L}_X}
-\frac{1}{4} \kappa  \mathcal{L}_X \phi'^2
-\frac{\mathcal{L}_X''}{2 \mathcal{L}_X}.
\end{eqnarray}
To canonicalize the quadratic action, we conduct~\cite{ZhongLiLiu2021} a field transformation
\begin{eqnarray}
 \label{EqFieldTrans}
G(t,r)\equiv \mathcal{L}_X^{1/2}\gamma^{1/4}\delta\phi,
\end{eqnarray}
 as well as a spatial coordinated transformation
 \begin{eqnarray}
 \label{EqcoorTrans}
r\to y\equiv \int dr \gamma^{-1/2},
\end{eqnarray}
after simplification, we obtain
 \begin{eqnarray}
\delta^{(2)}S=\frac12\int dt dy \left\{-\partial_{t}^2 G+\partial_{y}^2G-V_{\textrm{eff}}(y)G \right\} G,
\end{eqnarray}
where
 \begin{eqnarray}
V_{\textrm{eff}}(y)\equiv \frac{\partial_{y}^2 f}{f},\quad
f\equiv \mathcal{L}_X^{1/2}\gamma^{1/4}\frac{\partial_y \phi }{\partial_y A}.
\end{eqnarray}
Note that in order the definitions in Eqs.~\eqref{EqFieldTrans} and \eqref{EqcoorTrans} are meaningful, we demand
$\mathcal{L}_X> 0$ and $\gamma> 0$.

Obviously, the equation for the normal mode $G$ is
\begin{eqnarray}
-\partial_{t}^2 G+\partial_{y}^2G-V_{\textrm{eff}}(y)G=0.
\end{eqnarray}
After conducting the mode expansion
\begin{eqnarray}
G(t,y)=\sum_n e^{i \omega_n t}\psi_n(y),
\end{eqnarray}
one immediately sees that $\psi_n(y)$ satisfy a Schr\"odinger-like equation:
\begin{eqnarray}
\label{EqHamil}
\hat{H}\psi_n=-\frac{d^2\psi_n}{dy^2}+V_{\textrm{eff}}(y)\psi_n=\omega_n^2 \psi_n.
\end{eqnarray}
This equation is exactly the same one derived in Ref.~\cite{ZhongLiLiu2021}.

One can check that the Hamiltonian operator in Eq.~\eqref{EqHamil} can be factorized as follows\footnote{The definitions for $\hat{\mathcal{A}}$ and $\hat{\mathcal{A}}^{\dagger}$ here are different from those of Ref.~\cite{ZhongLiLiu2021}. The present definitions are consistent with those used in literatures of supersymmetric quantum mechanics~\cite{CooperKhareSukhatme1995}. }
\begin{eqnarray}
\hat{H}=\hat{\mathcal{A}} \hat{\mathcal{A}}^{\dagger} ,
\end{eqnarray}
where
\begin{eqnarray}
\hat{\mathcal{A}}=\frac{d}{d y}+\frac{\partial_{y} f}{f}, \quad \hat{\mathcal{A}}^{\dagger}=-\frac{d}{d y}+\frac{\partial_{y} f}{f} .
\end{eqnarray}
Usually, the factorization of the Hamiltonian operator indicate the spectrum is positive-semidefinite~\cite{InfeldHull1951}, and therefore the background kink solution is stable. The ground state is the zero mode with the following wave function:
\begin{eqnarray}
\psi_0\propto f.
\end{eqnarray}
Factorizable Hamiltonian operators are also closely related to the study of supersymmetric quantum mechanics~\cite{Witten1981,CooperKhareSukhatme1995}.
For example, the superpotential is simply
\begin{eqnarray}
W(y)= \frac{\partial_{y} f}{f},
\end{eqnarray}
in terms of which, the effective potential $V_{\textrm{eff}}$ can be written as
\begin{eqnarray}
V_{\textrm{eff}}(y)=W^2+\partial_y W.
\end{eqnarray}
Also, one can define a superpartner Hamiltonian for $\hat{H}$:
\begin{eqnarray}
\hat{H}_-=\hat{\mathcal{A}}^{\dagger} \hat{\mathcal{A}} ,
\end{eqnarray}
the corresponding potential is
\begin{eqnarray}
V_-(y) =f\partial_y^2(f^{-1})=W^2-\partial_y W.
\end{eqnarray}
When both $V_{\textrm{eff}}$ and $V_-$ are nonsingular, the spectra of  $\hat{H}$ and $\hat{H}_-$ are identical, except their zero modes. But if any one of the potentials is singular at some points, the supersymmetry might be broken, and the spectra are at most partially identical~\cite{CooperKhareSukhatme1995,JevickiRodrigues1984,RoyRoychoudhury1985,CasahorranNam1991,PanigrahiSukhatme1993}. Even worse, bound states with negative eigenvalues might appear in the spectrum of $\hat{H}_-$.

As shown in Refs.~\cite{Zhong2021,ZhongLiLiu2021}, the effective potential $V_{\textrm{eff}}$ for 2D symmetric gravitating kinks are typically divergent at $x=0=y$ (because $\partial_y A=0=\partial_x A$ here), and one must be careful with the stability issue.

\section{Conclusions}
\label{sec_con}
In this work, we noticed the fact that the effective potentials $V_{\textrm{eff}}$ for symmetric gravitating kinks are typically divergent at the origin, which means that the supersymmetry between $\hat{H}$ and $\hat{H}_-$ might be explicitly broken. If this is the case,  then not just the symmetry between the spectra of $\hat{H}$ and $\hat{H}_-$ are broken, the spectrum of $\hat{H}_-$ are not positive-semidefinite any more. On the other hand, the study of 5D thick branes has shown that if one starts with the equation of motion approach, but chooses an improper gauge condition (the longitude gauge, for instance), he might obtain $\hat{H}_-$ rather than $\hat{H}$. Only by starting with the quadratic action approach, could one obtain the correct Hamiltonian for the normal modes of perturbation.

For this reason, we reconsidered the stability issue of 2D gravitating kink solutions of Refs.~\cite{Zhong2021,ZhongLiLiu2021} from the quadratic action approach. Our results indicate that $\delta\varphi=0$ is a proper gauge condition for the present 2D gravitating kink model, in the sense that the Schr\"odinger-like equations obtained from the equation of motion approach and from the quadratic action approach are equivalent. We also showed that the present 2D model is so simple that all the calculations can be conducted with the help of some Mathematica codes. The result of the present work can also be applied to other static solutions.

\section*{Declaration of competing interest}
The authors declare that they have no known competing finan-cial interests or personal relationships that could have appeared to influence the work reported in this paper.
\section*{Acknowledgements}

This work was supported by the National Natural Science Foundation of China (Grant numbers~12175169, 11847211, 11605127), the Fundamental Research Funds for the Central Universities (Grant number~xzy012019052).




\end{document}